\documentclass[aps,superscriptaddress,,twocolumn,showpacs]{revtex4-1}
\pdfoutput=1
\usepackage{color}
\usepackage{amsmath, amsthm, amsfonts}    
\usepackage{amssymb}
\usepackage{mathptmx} 
\usepackage{dsfont}
\usepackage{enumitem} 
\usepackage{appendix}
\usepackage{cancel}
\usepackage{lipsum}
\usepackage{graphicx}
\usepackage{tikz}
\usepackage[unicode=true,pdfusetitle,
 bookmarks=true,bookmarksnumbered=false,bookmarksopen=false,
 breaklinks=true,pdfborder={0 0 0},backref=false,colorlinks=true,citecolor=blue]{hyperref}
\usepackage{graphicx}   
\usepackage[caption=false]{subfig}       
\usepackage{bm}            
\usepackage[normalem]{ulem}  

\newlength\imagewidth
\newlength\imagescale
\def\be{\begin{eqnarray}}
\def\ee{\end{eqnarray}}
\def\r{{\bf r}}

\def\E{{\bf E}}
\def\H{{\bf H}}

\def\F{{\bf F}}

\def\im{{\rm i}}

\definecolor{JOT-color}{named}{blue}
\definecolor{CSF-color}{named}{orange}

\begin{document}

\title{On the helicity conservation for Mie optical cavities}

\author{Jorge Olmos-Trigo}
\email{jolmostrigo@gmail.com}
\affiliation{Donostia International Physics Center (DIPC),  20018 Donostia-San Sebasti\'{a}n,  Spain}

\author{Xavier Zambrana-Puyalto}
\email{xavislow@protonmail.ch}
\affiliation{Istituto Italiano di Tecnologia, Via Morego 30, 16163 Genova, Italy.}

\begin{abstract}
The study of helicity in the context of light-matter interactions is an increasing area of research. However, some fundamental aspects of the helicity content of light fields inside scatterers have been overlooked. In this work, we demonstrate that the helicity of light fields inside lossless spherical cavities cannot be either conserved or sign-flipped as a result of a scattering event. The underlying reason is that the internal electric and magnetic Mie coefficients cannot simultaneously oscillate with equal amplitude and equal/opposite phase. Our analytical demonstration is fulfilled regardless of the refractive index, sphere size, and multipolar order. In addition, we show that the helicity of light fields inside lossy spheres can be conserved. This fact is in striking contrast to the behavior of the scattered field, whose helicity content cannot be conserved precisely when the sphere has losses. Finally, we show that the helicity content of internal fields can be flipped for materials with gain.
\end{abstract}

\maketitle

The study of light-matter interactions is ubiquitous across physical sciences. In particular, the increasing reach of Photonic technologies~\cite{Saleh1991} is due to the advances in controlling light-matter interactions.

Many different properties of light-matter interactions can be partially controlled or tweaked. In recent years, a property of light-matter interactions that had been overlooked for a long time has gained significant attention: electromagnetic helicity~\cite{Calkin1965}. The first fundamental explanation of the role of electromagnetic helicity in Maxwell equations dates back to 1965 when Calkin showed that the electromagnetic fields in vacuum satisfy a continuous symmetry: the electromagnetic duality. All continuous symmetries have a generator, and Calkin found helicity as the generator of duality for the electromagnetic fields in vacuum. Now, helicity is defined as the projection of the total angular momentum ($\bf{J}$) onto the linear momentum of the wave ($\bf{p}$), namely, $\Lambda =({\bf{J} \cdot \bf{p}}) /{|\bf{p}|}$~\cite{fernandez2012helicity}. Its definition becomes simpler in the plane wave decomposition of an electromagnetic field, as helicity is associated with the handedness of circular polarization of each plane wave to its momentum vector. However, due to the non-existence of magnetic monopoles in nature, which prevents Maxwell equations from being dual-symmetric in the presence of matter, the interest of the optical community in studying duality symmetry and helicity was residual for a large number of decades with very few exceptions~\cite{zwanziger1968quantum, bialynicki1994wave, bialynicki1981note}.

The study of helicity upon light-matter interactions took a clear boost in 2013 when it was theoretically
unveiled that helicity can be conserved upon scattering for dual materials, regardless of their geometry~\cite{fernandez2013electromagnetic}. A dual material is such that the ratio between its relative magnetic permeability and its electric permittivity happens to be constant ($\mu/\varepsilon = \text{constant}$). These materials restore duality symmetry in the macroscopic approximation of Maxwell equations. Unfortunately, there are no magnetic materials ($\mu \neq 1$) at optical frequencies. The implication is that it should be impossible to experimentally observe helicity conservation at optical frequencies. However, shortly after, it was demonstrated \cite{Zambrana-Puyalto2013} that high refractive index particles could approximately conserve helicity due to their electric and magnetic resonances~\cite{garcia2011strong, kuznetsov2016optically}. Since then, a lot of work has been done to characterize the role of helicity in light-matter interactions~\cite{Tischler2014,corbaton2016objects,alpeggiani2018electromagnetic,forbes2021measures}, as well as to study its relation with chirality~\cite{cameron2017chirality,gutsche2018optical,poulikakos2019optical, hanifeh2020optimally}. 

One of the first platforms that was used to study the role of helicity in light-matter interactions is Mie Theory~\cite{Mie1908}. Mie Theory studies the interaction of a plane wave with a spherical scatterer. It is a widely used platform to discover new analytical effects, as well as to predict experimental measurements~\cite{gouesbet2011generalized}. In particular, Mie Theory has been used to understand the role of helicity in a variety of scattering phenomena at the nanoscale, such as the so-called Kerker conditions~\cite{nieto2011angle, olmos2020optimal}, enhanced optical localization errors provided by optical mirages~\cite{Olmos-Trigo2019b}, or non-radiating sources~\cite{labate2017surface} such as hybrid optical anapoles~\cite{sanz2021multiple}. Moreover, helicity-dependent optical forces~\cite{liu2018three, nieto2021reactive} have given rise to a large plethora of interesting phenomena with applications in the biomedical and pharmaceutical industries, ranging from chiral sensing~\cite{graf2019achiral, lasa2020surface, D2MA00052K}  chiral sorting and enantioselective and enantiospecificdetection~\cite{hayat2015lateral, toyoda2013transfer, nieto2015opticaltheorem}, to optical tweezers~\cite{tkachenko2014helicity, ali2020enantioselective}, among others.

Most of the works dealing with helicity have studied scattering features, yet in the very last few years, some groups have started to study the helicity of fields trapped in a cavity~\cite{feis2020helicity, scott2020enhanced, doi:10.1021/acsphotonics.2c00134}. Having the enhancement of chiral sensing in mind, the authors of~\cite{feis2020helicity} came up with the design of a cavity that tightly concentrates light in areas where the helicity of the field is maintained. However, it is important to note that this intriguing and significant effect relies on numerical methods under very specific illumination conditions.
Here, we show that a simple lossy sphere can also be used to create a cavity that internally conserves the helicity content of the incident beam. In addition, we analytically demonstrate that the internal field of a lossless sphere cannot conserve the helicity content of an incoming beam. 
Note that our findings of the conservation of helicity for the internal field of a spherical cavity are in striking contrast with what is known about the scattered field, \textit{i.e.} the scattered field can only conserve the helicity of the incoming beam when the spherical cavity is lossless~\cite{olmos2020kerker}. Last but not least, we also show that the helicity of the internal field can be flipped if a sphere with optical gain is used.

Next, we lay out the framework that we will use to show how helicity works for internal fields. First, let us consider an incoming field with well-defined helicity $\sigma = \pm 1$~\cite{olmos2020unveiling,olmos2019sectoral},
\be
\E^{\sigma}_{\text{inc}} (k \r) &=&  \sum_{l=1}^{\infty} \sum_{m=-l}^{+l}  C_{lm}^{ \sigma} \boldsymbol{\Psi}_{lm}^{\sigma} (k \r).
\label{multipolar} 
\ee
Here $C_{lm}^{ \sigma}$ denotes the incoming coefficients characterizing the nature of the wave, $k$ is the radiation wavevector, and 
\be
\boldsymbol{\Psi}_{lm}^{\sigma} &=& \frac{1}{\sqrt{2}} \left[ {\boldsymbol{N}}_{lm} +  \sigma {\boldsymbol{M}}_{lm}  \right], \label{V1} \\
{\boldsymbol{M}}_{lm} &\equiv & j_l(kr)\boldsymbol{X}_{lm},  \quad
{\boldsymbol{N}}_{lm} \equiv \frac{1}{k} \boldsymbol{\nabla} \times {\boldsymbol{M}}_{lm}, \\
\boldsymbol{X}_{lm} &\equiv& \frac{1}{\sqrt{l(l+1)}} {\bf{L}} Y_{lm} (\theta,\varphi). \label{V3}
\ee
Here $\boldsymbol{M}_{lm}$ and $\boldsymbol{N}_{lm}$  are the so-called Hansen's multipoles~\cite{jackson1999electrodynamics}, $ j_l(kr)$ are the spherical Bessel functions,
$Y_{lm} (\theta,\varphi)$ are the spherical harmonics, $\theta$ and $\varphi$ being  the polar and azimuthal angles,  and $ {\bf{L}} =  \left\{ -\im \r \times \boldsymbol{\nabla}\right\} $ is the total angular momentum operator. 
Let us recall that the multipoles $ \boldsymbol{\Psi}_{lm}^{\sigma} $ 
are simultaneous eigenvectors of the squared angular momentum  $L^2$, the projection of the angular momentum on one direction $L_z$, and the helicity operator ${\Lambda}$~\cite{fernandez2013electromagnetic}, with eigenvalues $l(l+1)$, $m$ and $\sigma$, respectively~\cite{jackson1999electrodynamics}.
\begin{figure*}[t!]
    \centering
    \includegraphics[width= \textwidth]{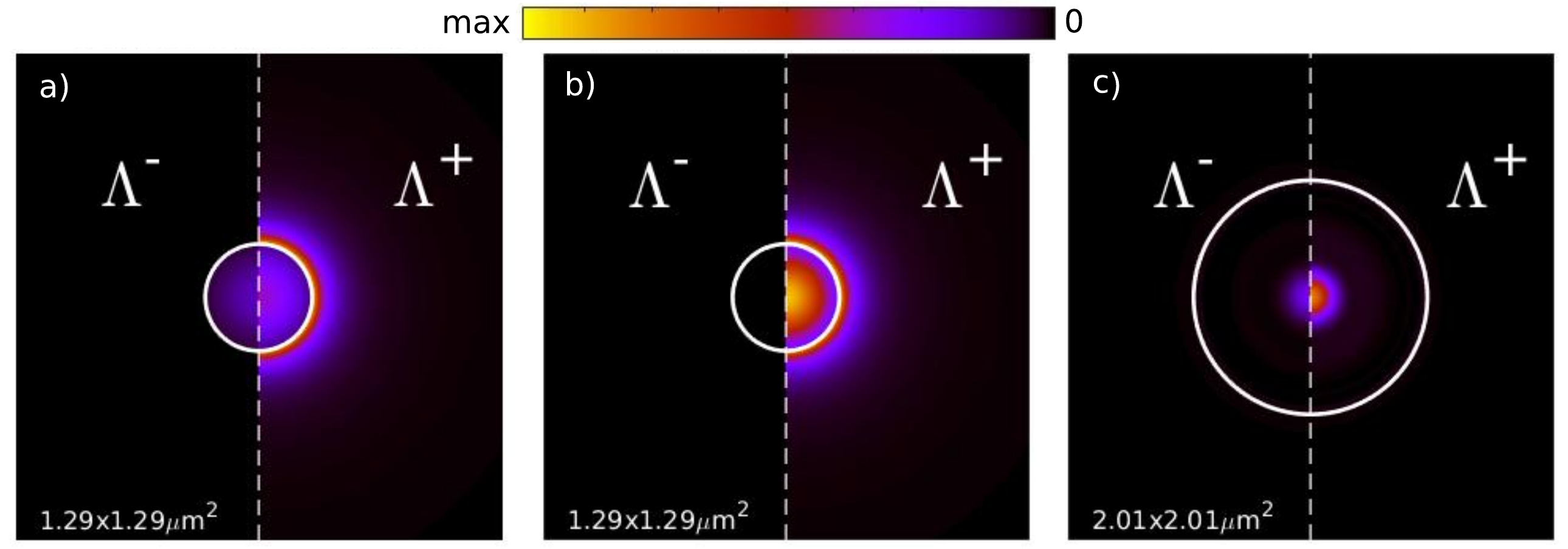}
    \caption{Decomposition of the internal and scattered fields  in the Riemann-Silberstein representation, $\Lambda^{\sigma'} = |\E + \im \sigma' Z \H|$, under illumination of a well-defined  helicity Gaussian beam ($\sigma = +1$). According to the colormap, black color represents the absence of the $\Lambda^{\sigma'}$ component. The white circle represents a lossless sphere with radius $a$ and refractive index contrast $m \in \mathbb{R}$.  a) Emergence of the First Kerker condition  provided by a Germanium (Ge)  sphere of $m = \sqrt{\epsilon} = 4.2$ and $a = 140$ nm  under illumination of a Gaussian beam focused with $\rm{NA}=0.15$ at $\lambda =1347$ nm. In this case, $|\Lambda_{\rm{sca}}^-| =0 $ while $|\Lambda_{\rm{int}}^-| \neq 0$. The plot has a dimension of $1.29 \times 1.29 \mu m^2$. b)  Electromagnetic duality restoration provided by a sphere of $\epsilon = \mu = 4.2$ and  $a = 140$ nm under illumination of a Gaussian beam focused with $\rm{NA}=0.15$ at $\lambda =1347$ nm. In this scenario,  $|\Lambda_{\rm{sca}}^-| = |\Lambda_{\rm{int}}^-| = 0$. The plot has a dimension of $1.29 \times 1.29 \mu m^2$. c) Emergence of the hybrid anapole mediated by a sphere of $m= \sqrt{\epsilon} = 3.33$ and $a = 453$ nm under the illumination of a Gaussian beam with $\rm{NA}=0.9$ at $\lambda =633$ nm. In this scenario, $|\Lambda_{\rm{sca}}^+| =|\Lambda_{\rm{sca}}^-| = 0$  while $|\Lambda_{\rm{int}}^-| \neq 0$ and $|\Lambda_{\rm{int}}^+| \neq 0$. The plot has a dimension of $2.01\times 2.01 \mu m^2$.}
    \label{F_1}
\end{figure*}
At this point, let us expand the electric field inside a sphere (internal electric field) in the same $ \boldsymbol{\Psi}_{lm}^{\sigma} $ basis. The internal field is obtained as the result of a scattering process involving the incoming field $\E^{\sigma}_{\text{inc}} (k \r) $ and a sphere of radius $a$:
\be\label{int}
\E^{\sigma}_{\rm{int}}(k_i \r) &=& \sum_{\sigma'=\pm1} \F^{\sigma \sigma'}_{\rm{int}}(k_i \r), \\ 
\F^{\sigma \sigma '}_{\rm{int}}(k_i \r) &=&   \sum_{l=0}^\infty \sum_{m=-l}^{+l} F^{\sigma \sigma' }_{lm}\boldsymbol{\Psi}_{lm}^{\sigma'}(k_i \r),\label{F_int}
\ee
where $F^{\sigma \sigma '}_{lm} =C^{\sigma}_{lm} \left( {d_l + \sigma \sigma' c_l}\right)/{2}$.
Here $d_l$ and $c_l$ denote the internal electric and magnetic Mie coefficients, respectively~\cite{bohren2008absorption} and $k_i = m k$,  with  $m = m_p/m_h$, $m_p$ and $m_h$  being the refractive index of the sphere and medium, respectively.

From Eq.~\eqref{F_int}, we can notice that  $d_l = c_l$ preserves  the incoming  helicity ($F^{\sigma \sigma'}_{lm} = C^{\sigma}_{lm} d_l \delta_{\sigma \sigma'}$). Similarly, it is straightforward to infer that $d_l = -c_l$ flips the incoming helicity ($F^{\sigma \sigma'}_{lm} = C^{\sigma}_{lm} d_l \left[1- \delta_{\sigma \sigma'} \right]$). 
At this stage, let us draw our attention to the  internal electric and magnetic Mie coefficients expressed in phase-shifts notation~\cite{luk2017hybrid},
\begin{align} \label{mie}
d_l =  -\frac{\im m}{F^{(a)}_l + \im G^{(a)}_l  }, && c_l =  \frac{\im m}{F^{(b)}_l + \im G^{(b)}_l  },
\end{align}
where
\be \label{Fa}
F^{(a)}_l &=& m \psi^{'}_{l} (q) \psi_{l}(mq) - \psi_{l}(q)\psi^{'}_{l}(mq),
\\ \label{Ga}
G^{(a)}_l &=& m x^{'}_{l}(q)\psi_{l}(mq) - \psi^{'}_{l}(mq)x_{l}(q),
 \\ \label{Fb}
F^{(b)}_l &=& m \psi^{'}_{l}(mq)\psi_{l}(q) - \psi_{l}(mq)\psi^{'}_{l}(q), \\
\label{Gb}
G^{(b)}_l &=& m x_{l}(q)\psi^{'}_{l}(mq) - \psi_{l}(mq)x^{'}_{l}(q).
\ee
Here $\psi_l(q) = (\frac{\pi q}{2})^{1/2} J_{l+\frac{1}{2}}(q)$ and $x_l(q) = (\frac{\pi q}{2})^{1/2} N_{l+\frac{1}{2}}(q)$ denote the Riccati-Bessel functions, where $J_l(q)$ and $N_l(q)$ are the Bessel functions of first and second kind, respectively~\cite{bohren2008absorption}; $'$ denotes the derivative with respect the argument,  $q = 2 \pi a / \lambda$ is the size parameter, and $\lambda$ is the radiation wavelength.
Now, we impose $d_l = \tau  c_l$, where $\tau = \pm 1$. After doing some algebra we arrive from Eq.~\eqref{mie}  to  
\be \label{phase}
F^{(a)}_l + \im G^{(a)}_l  = -\tau \left(F^{(b)}_l + \im G^{(b)}_l \right),
\ee
with
\begin{align} \label{F_AB}
F^{(a)}_l = m A - B, && G^{(a)}_l = m C - D,
\end{align}
\begin{align} \label{G_AB}
F^{(b)}_l = m B - A, && G^{(a)}_l = m D - C.
\end{align}
Here, we have defined $A = \psi'_l (q) \psi_{l}(mq)$, $B = \psi_{l}(q)\psi'_{l}(mq)$, $C = x'_{l}(q)\psi_{l}(mq)$, and $D = \psi'_{l}(mq)x_{l}(q)$. 
Now, by taking into account Eqs.\eqref{F_AB}-\eqref{G_AB}, it can be shown that $d_l = \tau c_l$, given by Eq.~\eqref{phase}, can be re-written as
\be \label{ABCD}
A + \tau B = -\im \left(C + \tau D \right).    
\ee
Now, by inspecting Eq.~\eqref{ABCD}, it is straightforward to re-write $d_l = \tau c_l$ as  
\begin{equation}
\tau  \psi'_{l}(mq) [\psi_{l}(q) + \im x_l(q) ] +  \psi_{l}(mq) [ \psi'_{l}(q) + \im x^{'}_l(q) ] = 0.
\end{equation}
At this stage and by identifying  $h_l(q) = \psi_{l}(q) + \im x_l(q)$ as the spherical Hankel function of second kind~\cite{bohren2008absorption}, we finally get
\begin{equation} \label{master}
 \psi_l(mq)h'_l(q) + \tau \psi'_l(mq)h_l(q) = 0.
\end{equation}
Eq.~\eqref{master} represents a notable simplification as it allows us to compute $d_l = \tau  c_l$  by making use of fundamental properties of just two spherical Bessel functions.
At this point, let us split the solutions of Eq.~\eqref{master} into two possible physical scenarios, namely,  non-absorbing ($m \in \mathbb{R}$) and lossy materials ($m \in \mathbb{C}$ with $|\Im \{m \}| \neq 0$). 
Let us first draw our attention to the lossless case ($m \in \mathbb{R}$), in which the complex Eq.~\eqref{master} can be re-expressed as two real-valued equations:
\begin{align} \label{F_ab}
&F^{(a)}_l = -\tau F^{(b)}_l \rightarrow  \psi_{l}(mq)\psi^{'}_{l}(q)  +\tau \psi^{'}_{l}(mq)\psi_{l}(q) = 0, \\ \label{G_ab}
&G^{(a)}_l = -\tau G^{(b)}_l  \rightarrow \psi_{l}(mq) x^{'}_{l}(q)+ \tau  \psi^{'}_{l}(mq)x_{l}(q) = 0.
\end{align}
Now, it is important to notice that Eqs.~\eqref{F_ab}-\eqref{G_ab} need to be simultaneously fulfilled to obtain  $d_l = \tau c_l$.
At this point, let us drive our attention to Lemma~\ref{Lemma_1}~\cite{watson1995treatise}: 
\begin{enumerate} 
\item  \label{Lemma_1} The zeros of any cylinder function or its derivative are
simple; If $\nu$ is real, then $J_\nu(z)$, $J'_\nu(z)$, $N_\nu(z)$, $N'_\nu(z)$ each
have an infinite number of positive real zeros. The $m$-th positive zeros of these spherical functions are denoted by $j_{\nu,m}$, $j'_{\nu,m}$, $n_{\nu,m}$, $n'_{\nu,m}$, and interlace according to $\nu<j'_{\nu,1}<n_{\nu,1}<n'_{\nu,1}<j_{\nu,1}<j'_{\nu,2}<n_{\nu,2}< ... \; .$
\end{enumerate}

Let us now use the Lemma~\ref{Lemma_1} to prove that the zeros of the Riccati-Bessel functions are also interlaced: Let be $\psi_\nu(z) = (\frac{\pi z}{2})^{1/2} J_{\nu+\frac{1}{2}}(z) = 0$. For $z  \neq 0$, this solution is given by the zeros of $J_{\nu+\frac{1}{2}}(z)$. Let us now take the first derivative of the Riccati-Bessel to then set it to zero, namely,
\begin{equation}\label{Riccati}
\psi'_\nu (z) = \frac{1}{2}{J_{\nu+\frac{1}{2}}(z)} + z J'_{\nu+\frac{1}{2}}(z) = 0.
\end{equation}
From Eq.~\eqref{Riccati}  and by taking into account that we have first settled $J_{\nu+\frac{1}{2}}(z) =0$, the only possible solution of $\psi'_\nu (z)  = 0$ might be given by $J'_{\nu+\frac{1}{2}}(z) = 0$. However, the latter cannot be met due to the interlaced property of the zeros of Bessel functions provided by Lemma~\ref{Lemma_1}. Consequently, the possible solutions of  Eqs.~\eqref{F_ab}-\eqref{G_ab} are notably reduced (see~\footnote{Notice that if $\psi_l'(z) = 0$ then $\psi_l(z) \neq 0$. On other hand,  if $\psi_l(z) = 0$ then $\psi'_l(z) \neq 0$. Notice that if $x_l'(z) = 0$ then $x_l(z) \neq 0$. On other hand,  if $x_l(z) = 0$ then $x'_l(z) \neq 0$.} to get insight into all possible combinations). That is, the only possibilities that make $d_l=\tau c_l$ when $m \in \mathbb{R}$ are: 
\begin{enumerate}[label=\roman*.]
    \item First Kerker condition (helicity conservation in scattering)~\cite{nieto2011angle,zambrana2013duality}, mathematically expressed as $\psi_l'(mq) =0$ (node of the first kind) and $\psi_l(mq) =0$ (node of the second kind)~\cite{olmos2020unveiling}.
    \item  Hybrid or Kerker anapoles (optical transparency)~\cite{luk2017hybrid}, mathematically given by $\psi_l'(mq) =0$ and $\psi_l'(q) =0$ or $\psi_l(mq) =0$ and $\psi_l(q) =0$~\cite{sanz2021multiple}.
\end{enumerate}


First, the conservation of helicity in scattering for a lossless sphere precludes the conservation of  the internal helicity   (see Fig~\ref{F_1}a)). Mathematically, we can demonstrate this by using the property that the zeros of the Riccati-Bessel functions are interlaced. That is, it is impossible to fulfill both Eqs~\eqref{F_ab}-\eqref{G_ab} if $\psi_l'(mq) =0$ or $\psi_l(mq) =0$. On physics grounds, this is a result of the fact that we are dealing with spheres with $\mu =1$ and $\epsilon^2=m \in \mathbb{R}$, with $\epsilon$ and $\mu$ denoting the electric permittivity and magnetic permeability, respectively. Now, it is known that electromagnetic duality can only be restored when $\epsilon = \mu$~\cite{fernandez2013electromagnetic}. When duality is restored, helicity is conserved for both the internal and scattered field, as it can be observed in Fig.~\ref{F_1}b). However, for non-magnetic lossless particles, this physical picture is precluded: if the scattered helicity is preserved, then the internal helicity is not. 

Second, hybrid anapoles, namely spectral points in which the scattering associated with a given order $l$ vanishes, also prevent $d_l = \tau c_l$. Mathematically, the hybrid anapole condition imposes that $F^{(a)}_l = F^{(b)}_l = 0$, which yields $c_l / d_l = m$ when $\psi_l'(mq) =0$ and $\psi_l'(q) =0$, and  $d_l / c_l = m$ when $\psi_l(mq) =0$ and $\psi_l(q) =0$~\cite{sanz2021multiple}. Physically, we can notice that hybrid anapoles are a particular solution of the first Kerker condition with zero scattering, and as explained above: a sphere with $\mu=1$ cannot be dual. As a result, we conclude that hybrid anapoles of order $l$ not only inhibit the conservation of the internal helicity, as can be inferred from Fig~\ref{F_1}c), but also constrain the internal Mie coefficients to have a very specific relation given by  $c_l / d_l = m^{\pm1}$.
Remarkably, this phenomenon remains valid regardless of the refractive index, optical size, making our proof general for anapoles associated with higher multipolar orders.

\begin{figure*}[t!]
    \centering
    \includegraphics[width= \textwidth]{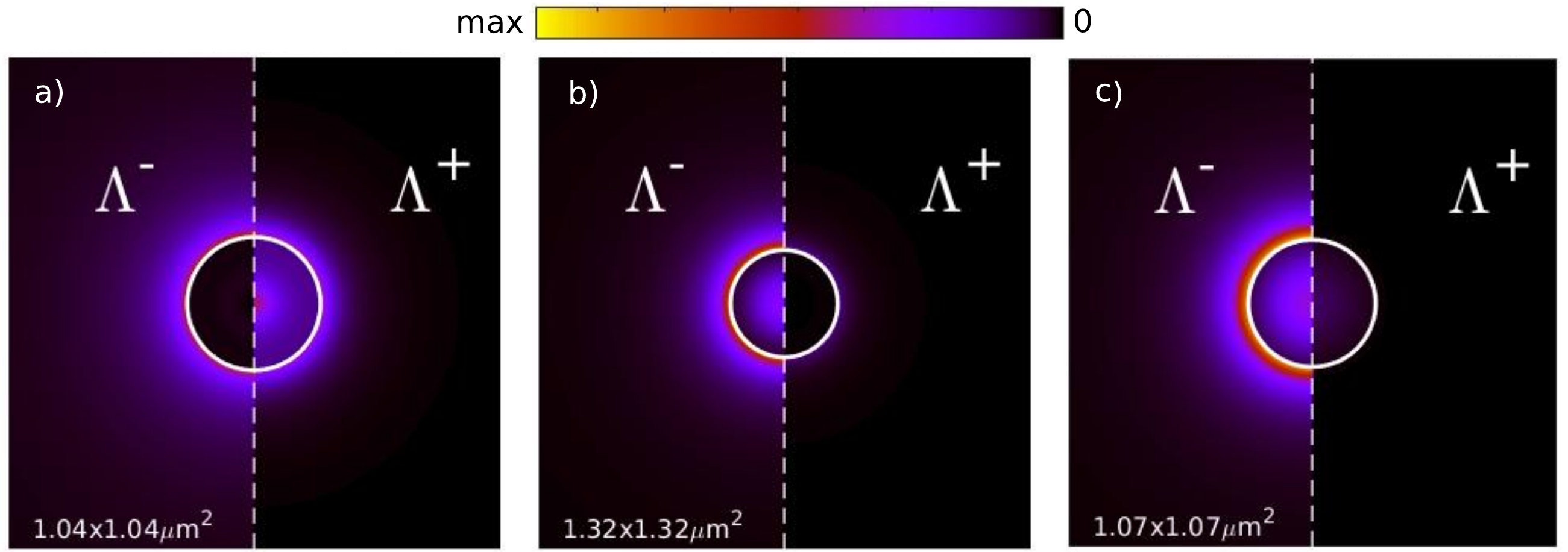}
    \caption{Decomposition of the internal and scattered fields  in the Riemann-Silberstein representation, $\Lambda^{\sigma'} = |\E + \im \sigma' Z \H|$, under illumination of a well-defined  helicity Gaussian beam ($\sigma = +1$). According to the colormap, black color represents the absence of the $\Lambda^{\sigma'}$ component. The white circle represents the lossy sphere with radius $a$ and refractive index contrast $m \in \mathbb{C}$. a) Conservation of the helicity content of light fields inside a sphere of $m=4.2 + 0.54 \im $ and $a = 140$ nm under illumination of a Gaussian beam with $\rm{NA}=0.15$ at $\lambda =407$ nm. In this scenario,  the scattered helicity is not preserved while $|\Lambda_{\rm{int}}^-| = 0$. The plot has a dimension of $1.04\times 1.04 \mu m^2$. b) Flipping of the helicity content of light fields  provided by a sphere of $m = \sqrt{\epsilon} = 4.2 -0.37 \im$ and $a = 140$ nm  under illumination of a Gaussian beam with {$\rm{NA}=0.15$} at $\lambda =1038$ nm. In this case, $|\Lambda_{\rm{int}}^+| =0$. The plot has a dimension of $1.32\times 1.32 \mu m^2$. c) Second Kerker condition  provided by a sphere of $m = \sqrt{\epsilon} = 4.2 -0.33 \im$ and $a = 140$ nm  under illumination of a Gaussian beam with $\rm{NA}=0.15$ at $\lambda =1038$ nm. In this case, $|\Lambda_{\rm{sca}}^+| =0$. The plot has a dimension of $1.07 \times 1.07 \mu m^2$.}
    \label{F_2}
\end{figure*}


Hitherto, we have found that the conservation of helicity in scattering and the emergence of hybrid anapoles are related to the fact that the helicity of light fields inside a lossless sphere cannot be either conserved or sign-flipped as a result of a scattering event; namely, $d_l \neq \tau c_l$ for $m \in \mathbb{R}$. In this vein, it has been analytically proven in Ref.~\cite{olmos2020kerker} and Ref.~\cite{sanz2021multiple} that the conservation of helicity in scattering and the emergence of hybrid anapoles can only occur for lossless particles. Next, we show that conservation of helicity for the internal fields can only happen for the opposite case, \textit{i.e.} for spheres made of a lossy material where $m \in \mathbb{C}$ with $|\Im \{m \}| \neq 0$. To show that, let us consider $d_l = \tau c_l$ without assuming constraints on the real and imaginary parts in Eq.~\eqref{master}. 
Taking this crucial fact into account, we can re-write $d_l = \tau c_l$ as
\begin{equation} \label{trascendental}
-\tau =   \frac{\psi_l(mq)h'_l(q) }{\psi'_l(mq)h_l(q) }.
\end{equation}
Eq.~\eqref{trascendental} is a transcendental equation that can only be satisfied for $|\Im \{m \}| \neq 0$. As a matter of fact, it can be shown that for $\tau = +1$ the  set of solutions can only be given for   $\Im \{m \} > 0$ while for $\tau = -1$ these can only be met for   $\Im \{m \} < 0$. This phenomenon is  depicted in Fig.~\ref{F_2}. In particular, in Fig.~\ref{F_2}a), we show a Ge-like nanosphere in the visible spectral range ($\lambda = 407 $ nm, $m= 4.2 +0.54 \im$) whose internal field conserves the helicity content of the incident beam. In Fig.~\ref{F_2} we also show two spheres in the presence of optical gain. On the one hand, in Fig.~\ref{F_2}b) we show the case of a sphere where the helicity content of the internal field is just the opposite with respect to the incoming beam. On the other hand, in Fig.~\ref{F_2}c) we show a sphere whose scattered field helicity content is just the opposite of the incident beam. 

In conclusion, in our work we have unraveled a fundamental property of the internal Mie coefficients: we have demonstrated that the helicity of light fields inside lossless spheres cannot be either conserved or sign-flipped as a result of a scattering event. This demonstration is analytical and involves special features of the scattered field, such as the conservation of helicity and the emergence of hybrid anapoles. Our proof does not depend on the incident polarization, optical size, and multipolar order. We have also shown that, in striking contrast to the behavior of the scattered field, losses are a compulsory requirement to conserve the helicity content of the internal field. That is, $d_l=c_l$ can only happen for lossy materials. Finally, we have shown that the helicity content of internal fields can be flipped for materials with gain. Note that the main findings of helicity conservation inside spherical cavities  can be extrapolated to other geometries such as disks or cylinders, finding potential applications in chiral sensing and chiral spectroscopy techniques.

J. O. T. acknowledges support from the Spanish Ministerio de Ciencia e Innovacion (PID2019-
109905GA-C2) and from Eusko Jaurlaritza (KK-2021/00082). Moreover, J. O. T. acknowledges funding from the Basque Government’s IKUR initiative on Quantum technologies (Department of Education).

\bibliography{Bib_tesis}

\end{document}